# Limits on the global Hubble constant and the age of the universe from the local Hubble constant measurement

Yasushi Suto, Tatsushi Suginohara, and Yuichiro Inagaki

*Department of Physics, The University of Tokyo, Bunkyo-ku, Tokyo 113, Japan*

While the recent discovery of the Cepheid variables in the Virgo cluster galaxies puts additional support for the Hubble constant $H_0 \sim 80$km/sec/Mpc, a relatively lower value $H_0 \sim 50$km/sec/Mpc is suggested by other distance indicators based on the Sunyaev-Zel'dovich effect and the gravitational lens which probe the universe at higher redshifts $z = (0.1 \sim 1)$. In order to reconcile the possible discrepancy between the estimates of the Hubble constants from nearby galaxy samples and high-redshift clusters, we consider a model of locally open universe embedded in the spatially flat universe. We find analytic expressions for the lower limit on the global Hubble constant $H_G$, and the upper limit on the age of the universe with a given value for the Hubble constant $H_L$ in the local universe. We conclude that it is quite unlikely that the above difference in the estimates of the Hubble constant is explained within the framework of the gravitational instability picture.

Recent discovery[1, 2] of the Cepheid variables in the Virgo cluster galaxies strongly supports that the Hubble constant $H_0$, at least in our local ( $\lesssim 100h^{-1}$Mpc ) universe is close to 80 km/sec/Mpc, where $h$ is $H_0$ in units of 100 km/sec/Mpc; to be more specific, $h = 0.87 \pm 0.07$ from the NGC4517 distance[1] and $0.80 \pm 0.17$ from the M100 distance[2]. In fact, these values with the quoted errors are in good agreement with those suggested by previous observations based on independent distance indicators mainly in optical wavebands (see Ref.[3] for a review).

On the other hand, other distance indicators probing somewhat higher redshifts often suggest a systematically smaller value around $H_0 = 50$ km/sec/Mpc. In particular several observations on the basis of the Sunyaev-Zel'dovich (SZ) effect yield, for instance, $h = 0.65 \pm 0.25$ for A2218($z = 0.171$)[4], $0.47 \pm 0.17$ for A665 ($z = 0.182$)[5], and $0.41^{+0.15}_{-0.12}$ for CL0016+16 ($z = 0.545$)[5]. If we take the above estimates seriously, they are consistent, within the quoted error, only if $H_0 \sim 50$km/sec/Mpc, which is systematically smaller than the values inferred from observations of our local universe.

This allows two simple interpretations; (i) the estimator on the basis of the SZ effect is affected by some systematic underestimate bias, or (ii) the expansion rate in our local universe is larger than the global average. It is true that various possible sources may effect the estimate of $H_0$ using the Sunyaev-Zel'dovich effect. According to our recent quantitative re-examination[8], we showed that several factors, in particular the finite extension of X-ray clusters and the temperature profile, would result in $\sim 10$ percent uncertainty in $H_0$, but did not find any strong underestimate bias to the extent of which could account for the reported discrepancy. In fact, the recent estimate[9] of $H_0$ through the SZ effect of the Coma cluster ($z = 0.0235$) derives $H_0 = (74\pm29)$km/sec/Mpc, completely consistent with the other optical methods. Incidentally an analysis of the gravitational lensing of QSO 0957+561 ($z = 1.41$; supposedly due to the lensing galaxy at $z = 0.36$) indicates $h = 0.50 \pm 0.17$ [10], and $0.11 < h < 0.69$[11]. Although $H_0$ estimated from the gravitational lensing effect is generally regarded to be less convincing due to its sensitivity to the assumed model[6, 7], it is interesting to note that the derived values are consistent with the SZ estimator.

In the light of this, the second interpretation deserves a detailed theoretical study. A natural idea to account for the different expansion rates of our local ($r \lesssim 100h^{-1}$Mpc) and global ($z \gtrsim 0.2$) universes is to assume that our local universe is underdense compared with the universe on average. To be more specific, we consider a local universe with the density parameter $\Omega_L$, the local Hubble constant $H_L$, and the dimensionless cosmological constant $\lambda_L \equiv \Lambda/(3H_L^2)$ embedded in the spatially flat universe with $(\Omega_G, H_G, \lambda_G)$. A non-vanishing cosmological constant is suggested, at least in the global universe, from the age of the globular cluster $t_{gl} = 17 \pm 2$ Gyr[12], and also from the structure formation[13, 14].

In this Letter, we examine the extent to which this simplest model can explain the observation, and derive formulae of the lower limit on the global Hubble constant $H_G$, and of the upper limit on the age of the universe $t_0$, with a given value for $H_L$. Turner, Cen and Ostriker[15] computed the probability distribution function of the $H_L$ in $\Omega_G = 1.0$ and $\lambda_G = 0$ cold dark matter universe using numerical simulations with a given $H_G$. Our model deals with the problem analytically assuming the spherically symmetric local universe implicitly, and also considers general spatially flat models with non-vanishing $\lambda_G$.

We assume that the global universe is spatially flat:

$$\Omega_G \leq 1, \qquad \Omega_G + \lambda_G = 1. \tag{1}$$

Since the dimensional cosmological constant $\Lambda$ is most naturally supposed to be the same everywhere, $\lambda_L$ is related to $r \equiv (H_G/H_L)$ and $\Omega_G$ as

$$\lambda_L = r^2 \lambda_G = r^2(1 - \Omega_G). \tag{2}$$

So if the difference of $H_L$ and $H_G$ results from the purely gravitational evolution, the two universes should have the same age derived from the usual cosmic expansion law[14, 16]:

$$t_0(\Omega_G, \lambda_G, H_G) = t_0(\Omega_L, \lambda_L, H_L), \tag{3}$$

where

$$t_0(\Omega_G, \lambda_G, H_G) = \frac{1}{3H_G\sqrt{1-\Omega_G}} \ln \frac{2 - \Omega_G + 2\sqrt{1-\Omega_G}}{\Omega_G}, \tag{4}$$

and

$$t_0(\Omega_L, \lambda_L, H_L) = \frac{1}{H_L} \int_0^1 \frac{a\, da}{\sqrt{\Omega_L a + (1 - \Omega_L - \lambda_L)a^2 + \lambda_L a^4}}. \tag{5}$$

For a later convenience, let us define

$$\tau_L \equiv H_L t_0, \tag{6}$$

and

$$y = y(\Omega_G) \equiv \left(\frac{2 - \Omega_G + 2\sqrt{1-\Omega_G}}{\Omega_G}\right)^{1/3}. \tag{7}$$

Then eq.(3) is rewritten as

$$\tau_L = \frac{1}{r\sqrt{1-\Omega_G}} \ln y(\Omega_G)$$

$$= \frac{1}{r} \int_0^1 \frac{a\, da}{\sqrt{\Omega_L(a - a^2)/r^2 + a^2/r^2 + (1 - \Omega_G)(a^4 - a^2)}}. \tag{8}$$

In the simplest case of $\Omega_G = 1$ ($\lambda_G = 0$), one can explicitly solve eq.(8) for $r = r(\Omega_L, \Omega_G)$:

$$\frac{2}{3r(\Omega_L, \Omega_G = 1)} = \frac{1}{1 - \Omega_L} - \frac{\Omega_L}{2(1 - \Omega_L)^{3/2}} \ln \frac{2 - \Omega_L + 2\sqrt{1 - \Omega_L}}{\Omega_L}. \tag{9}$$

Therefore it is easy to see $r(\Omega_L, \Omega_G = 1) \geq r(\Omega_L = 0, \Omega_G = 1) = 2/3$. While eq.(8) cannot be solved for $r = r(\Omega_L, \Omega_G)$ explicitly unless $\Omega_G = 1$, the lower limit for $r$ can be obtained analytically:

$$r(\Omega_L, \Omega_G) \geq r(\Omega_L = 0, \Omega_G) = \frac{y^2 - 1}{y^2 + 1} \frac{y^3 + 1}{y^3 - 1}. \tag{10}$$

For $0 < \Omega_G < 1$ ($\infty > y > 1$), $r(\Omega_L = 0, \Omega_G)$ decreases monotonically from 1 to 2/3. In other words, as a non-vanishing cosmological constant increases, the lower bound for $H_G$ also increases to match the given value of $H_L$. Using the relation (10), one immediately sees that

$$\tau_L(\Omega_L, \Omega_G) \equiv H_L t_0(\Omega_L, \Omega_G) = \frac{\ln y(\Omega_G)}{r(\Omega_L, \Omega_G)\sqrt{1 - \Omega_G}} \leq \frac{y^2 + 1}{y^2 - 1} \ln y. \tag{11}$$

The limits on $r$ and $\tau_L$ (eqs.[10] and [11]) are summarized in Table 1 for several different values of $\Omega_G$. For comparison, we list the age of the universe in the case of $\Omega_L = \Omega_G$ and $\lambda_L = 1 - \Omega_G$ with the same $H_L$ in the fifth column, and the last column is the ratio of $t_0(\Omega_L = 0, \lambda_L = r^2 - r^2\Omega_G)$ and $t_0(\Omega_L = \Omega_G, \lambda_L = 1 - \Omega_G)$. In order to construct specific examples of the locally open universes in the spatially flat universe, we numerically solve eq.(8) to find the $\Omega_L - r$ relation for fixed $\Omega_G$ and $\tau_L$ (Fig.1). One clearly sees that the $r$ could be significantly smaller than unity only when $\Omega_G$ is close to unity and thus $\tau_L$ is small. If we adopt $t_0 = 17 \pm 2 \text{Gyr}$, or equivalently $\tau_L = (1.39 \pm 0.16)(0.8/h_L)$, for instance, $r$ cannot be less than $\sim 0.8$ even if $\Omega_L = 0$. If $\Omega_L \gtrsim 0.05$ in turn, $\tau_L \gtrsim 1.2$ implies that $r$ should be between 0.9 and 1. Therefore it is unlikely that realistic models with $\Omega_L \gtrsim 0.05$ could account for the possible difference of $H_G$ and $H_L$ described above.

Fig.2 shows the contour-lines for the ratio of local to global densities, i.e., $\rho_L/\rho_G = \Omega_L/(r^2\Omega_G)$ on the $\Omega_L - r$ plane. As expected, the value of $r$ is close to unity unless our local universe is significantly underdense compared with the global mean. If the density field smoothed over the scale of our local universe (around $100h^{-1}\text{Mpc}$) is random-Gaussian, the probability that $\Omega_L$ is less than a threshold $\Omega_c$ is given by

$$\text{Prob}(\Omega_L < \Omega_c) = \frac{1}{\sqrt{2\pi\sigma^2}} \int_{-\infty}^{\Omega_c/(r^2\Omega_0)-1} \exp\left(-\frac{\delta^2}{2\sigma^2}\right) d\delta = \frac{1}{2} \, \text{erfc}\left(\frac{1}{\sqrt{2}\sigma} - \frac{\Omega_c}{\sqrt{2}\sigma r^2\Omega_G}\right), \tag{12}$$

where $\sigma$ is the rms values of density fluctuations smoothed over the effective size of the local universe, and erfc($x$) is the complementary error function. For comparison, $\sigma(100h^{-1}{\rm Mpc})$ in the spatially flat cold dark matter models with $h = 0.5$ are 0.16, 0.11, and 0.034 for $\Omega_G = 0.1$, 0.2 and 1.0, respectively, where the amplitude is normalized so that $\sigma(8h^{-1}{\rm Mpc}) = 1$. The corresponding probabilities for $\rho_L/\rho_G < 0.3$ are computed with eq.(12) yielding $6 \times 10^{-6}$, $2 \times 10^{-11}$, and $10^{-95}$ again for $\Omega_G = 0.1$, 0.2 and 1.0, respectively. Furthermore the probabilities are significantly reduced for smaller $\rho_L/\rho_G$. Just for illustration, let us accept $H_L = 80 \pm 17$km/sec/Mpc from the M100 Cepheid and $H_G = 47 \pm 17$km/sec/Mpc from the CL 0016+16 SZ effect. Then $r = H_G/H_L$ becomes $0.59 \pm 0.34$, which is not easy to be accounted for, even with the quoted error, in the standard model of the structure formation. The conclusion, however, is crucially dependent on the choice of the effective volume of the local universe we consider, and also on the fluctuation amplitude (or the adopted cosmological models[16]) over the corresponding scales.

We have considered a model of locally open universe embedded in the spatially flat universe. Although no spatially flat models satisfy, for instance, $H_G = 80$km/sec/Mpc, $t_0 \geq 17$Gyr, and $\Omega_G \geq 0.1$ simultaneously, they can be reconciled in the present model due to the fact that the expansion rate in our local universe differs from the global average. We have shown, however, that matching the two universes places rather restrictive limits on $H_G/H_L$ and $\tau_L$ as long as purely gravitational evolution is responsible for the difference in $H_G$ and $H_L$. If the difference between $H_G$ and $H_L$ as tentatively suggested in some current observations is real, it would seriously challenge standard gravitational instability picture of the structure formation in the universe[14, 16].

This research was supported in part by the Grants-in-Aid by the Ministry of Education, Science and Culture of Japan (05640312, 06233209, 06740183).

**Table 1.** Limits on $r \equiv H_G/H_L$ and $\tau_L \equiv H_L t_0$.

| $\Omega_G$ | $y$ | $r(\Omega_L = 0)$ | $\tau_L(\Omega_L = 0)$ | $\tau_L(\Omega_L = \Omega_G, \lambda_L = 1 - \Omega_G)$ | ratio |
|---|---|---|---|---|---|
| 0.05 | 4.3 | 0.92 | 1.62 | 1.49 | 1.09 |
| 0.1 | 3.4 | 0.88 | 1.45 | 1.28 | 1.13 |
| 0.2 | 2.6 | 0.83 | 1.29 | 1.08 | 1.20 |
| 0.3 | 2.2 | 0.80 | 1.21 | 0.96 | 1.25 |
| 0.4 | 2.0 | 0.77 | 1.15 | 0.89 | 1.30 |
| 0.6 | 1.6 | 0.73 | 1.08 | 0.79 | 1.38 |
| 0.8 | 1.4 | 0.69 | 1.03 | 0.72 | 1.44 |
| 1.0 | 1.0 | 2/3 | 1.00 | 2/3 | 1.50 |

# Figure captions

**Figure 1.** Contour-lines for global density parameter $\Omega_G$ and the age of the universe $\tau_L$ on the $\Omega_L - r$ plane. Solid curves correspond to $\Omega_G = 0.05, 0.1, 0.2, 0.5$, and $1.0$, while dotted curves to $\tau_L = H_L t_0 = 1.0, 1.2, 1.4, 1.6$, and $1.8$. Note that $t_0 \sim 12.2\tau_L(0.8/h_L)$Gyr.

**Figure 2.** Density contour-lines $\rho_L/\rho_G = \Omega_L/(r^2\Omega_G)$ on the $\Omega_L - r$ plane. The curves correspond to $\rho_L/\rho_G = 0.1, 0.2, 0.3, 0.4, 0.5, 0.6, 0.7, 0.8$, and $0.9$, and are terminated where $\Omega_G$ exceeds unity.